\documentclass[twocolumn,preprintnumbers,secnumarabic,showpacs,amsmath,amssymb]{revtex4}
\usepackage{docs}

\usepackage{graphicx}
\usepackage{dcolumn}

\usepackage{bm}

\begin{document}
\title{Non-existence of static, spherically symmetric
and stationary, axisymmetric traversable wormholes coupled to
nonlinear electrodynamics}%


\author{Aar\'{o}n V. B. Arellano}%
\email{vynzds@yahoo.com.mx} \affiliation{Facultad de Ciencias,
Universidad Autonoma del Estado de M\'{e}xico, \\
El Cerrillo, Piedras Blancas, C.P. 50200, Toluca, M\'{e}xico \\
and Centro de Astronomia
e Astrof\'{\i}sica da Universidade de Lisboa,\\
Campo Grande, Ed. C8 1749-016 Lisboa, Portugal}

\author{Francisco S. N. Lobo}%
\email{flobo@cosmo.fis.fc.ul.pt} \affiliation{Centro de Astronomia
e Astrof\'{\i}sica da Universidade de Lisboa,\\
Campo Grande, Ed. C8 1749-016 Lisboa, Portugal}

\begin{abstract}

In this work we explore the possible existence of static,
spherically symmetric and stationary, axisymmetric traversable
wormholes coupled to nonlinear electrodynamics. Considering static
and spherically symmetric $(2+1)$ and $(3+1)-$dimensional wormhole
spacetimes, we verify the presence of an event horizon and the
non-violation of the null energy condition at the throat. For the
former spacetime, the principle of finiteness is imposed, in order
to obtain regular physical fields at the throat. Next, we analyze
the $(2+1)-$dimensional stationary and axisymmetric wormhole, and
also verify the presence of an event horizon, rendering the
geometry non-traversable. Relatively to the $(3+1)-$dimensional
stationary and axisymmetric wormhole geometry, we find that the
field equations impose specific conditions that are incompatible
with the properties of wormholes. Thus, we prove the non-existence
of the general class of traversable wormhole solutions, outlined
above, within the context of nonlinear electrodynamics.

\end{abstract}

\pacs{04.20.Jb, 04.40.Nr, 11.10.Lm}

\maketitle

\section{Introduction}

A specific model of nonlinear electrodynamics was proposed by Born
and Infeld in 1934 \cite{BI} founded on a principle of finiteness,
namely, that a satisfactory theory should avoid physical
quantities becoming infinite. The Born-Infeld model was inspired
mainly to remedy the fact that the standard picture of a point
particle possesses an infinite self-energy, and consisted on
placing an upper limit on the electric field strength and
considering a finite electron radius.
Later, Pleba\'{n}ski presented other examples of nonlinear
electrodynamic Lagrangians \cite{Pleb}, and showed that the
Born-Infeld theory satisfies physically acceptable requirements. A
further discussion of these properties can be found in Ref.
\cite{Birula}.
Furthermore, a recent revival of nonlinear electrodynamics has
been verified, mainly due to the fact that these theories appear
as effective theories at different levels of string/M-theory, in
particular, in D$p-$branes and supersymmetric extensions, and
non-Abelian generalizations (see Ref. \cite{Witten} for a review).

Much interest in nonlinear electrodynamic theories has also been
aroused in applications to cosmological models, in particular, in
explaining the inflationary epoch and the late accelerated
expansion of the universe \cite{Novello,Moniz}. In this
cosmological context, an inhomogeneous and anisotropic nonsingular
model for the universe, with a Born-Infeld field was studied
\cite{Sal-Breton}, the effects produced by nonlinear
electrodynamics in spacetimes conformal to Bianchi metrics were
further analyzed \cite{Sal-Breton2}, and geodesically complete
Bianchi spaces were also found \cite{Sal-Breton3}. Homogeneous and
isotropic cosmological solutions governed by the non-abelian
Born-Infeld Lagrangian \cite{DGZZ}, and anisotropic cosmological
spacetimes, in the presence of a positive cosmological constant
\cite{Vollick}, were also extensively analyzed.
In fact, it is interesting to note that the first {\it exact}
regular black hole solution in general relativity was found within
nonlinear electrodynamics \cite{Garcia,Garcia2}, where the source
is a nonlinear electrodynamic field satisfying the weak energy
condition, and the Maxwell field is reproduced in the weak limit.
It was also shown that general relativity coupled to nonlinear
electrodynamics leads to regular magnetic black holes and
monopoles \cite{Bronnikov1}, and regular electrically charged
structures, possessing a regular de Sitter center
\cite{Dymnikova}, and the respective stability of these solutions
was further explored in Ref. \cite{Breton-BH}.

Recently, an alternative model to black holes was proposed, in
particular, the gravastar picture \cite{gravastar}, where there is
an effective phase transition at or near where the event horizon
is expected to form, and the interior is replaced by a de Sitter
condensate. The gravastar model has no singularity at the origin
and no event horizon, as its rigid surface is located at a radius
slightly greater than the Schwarzschild radius. In this context, a
gravastar model within nonlinear electrodynamics, where the
interior de Sitter solution is substituted with a Born-Infeld
Lagrangian, was also found. This solution was denoted as a
Born-Infeld phantom gravastar \cite{Bilic}.

Relatively to wormhole spacetimes \cite{Morris,Visser}, an
important and intriguing challenge is the quest to find a
realistic matter source that will support these exotic geometries.
The latter are supported by {\it exotic matter}, involving a
stress energy tensor that violates the null energy condition
(NEC), i.e., $T_{\mu\nu}k^\mu k^\nu \geq 0$, where $T_{\mu\nu}$ is
the stress-energy tensor and $k^{\mu}$ any null vector. Several
candidates have been proposed in the literature, for instance, to
cite a few, null energy condition violating massless conformally
coupled scalar fields supporting self-consistent classical
wormholes \cite{barcelovisserPLB99}; the extension of the
Morris-Thorne wormhole with the inclusion of a cosmological
constant \cite{LLQ-PRD}; and more recently, the theoretical
realization that wormholes may be supported by exotic cosmic
fluids, responsible for the accelerated expansion of the universe,
such as phantom energy \cite{phantomWH} and the generalized
Chaplygin gas \cite{ChaplyginWH}. It is also interesting to note
that an effective wormhole geometry for an electromagnetic wave
can appear as a result of the nonlinear character of the field
\cite{Novello2}.

In Ref \cite{Arell-Lobo}, evolving $(2+1)$ and $(3+1)-$dimensional
wormhole spacetimes, conformally related to the respective static
geometries, within the context of nonlinear electrodynamics were
also explored. It was found that for the specific
$(3+1)-$dimensional spacetime, the Einstein field equation imposes
a contracting wormhole solution and the obedience of the weak
energy condition. Furthermore, in the presence of an electric
field, the latter presents a singularity at the throat. However, a
regular solution was found for a pure magnetic field. For the
$(2+1)-$dimensional case, it was also found that the physical
fields are singular at the throat. Thus, taking into account the
principle of finiteness, that a satisfactory theory should avoid
physical quantities becoming infinite, one may rule out evolving
$(3+1)-$dimensional wormhole solutions, in the presence of an
electric field, and the $(2+1)-$dimensional case coupled to
nonlinear electrodynamics.

In this work we shall be interested in exploring the possibility
that nonlinear electrodynamics may support static, spherically
symmetric and stationary, axisymmetric traversable wormhole
geometries. In fact, Bronnikov \cite{Bronnikov1,Bronnikov2} showed
that static and spherically symmetric $(3+1)-$dimensional
wormholes is not the case, and we shall briefly reproduce and
confirm this result. We further consider the $(2+1)-$dimensional
case, which proves to be extremely interesting, as the principle
of finiteness is imposed, in order to obtain regular physical
fields at the throat. We shall next analyze the $(2+1)$ and
$(3+1)-$dimensional stationary and axisymmetric case \cite{Teo}
coupled to nonlinear electrodynamics.

This paper is outlined in the following manner: In Sec.
\ref{StaticWH} we analyze $(2+1)$ and $(3+1)-$dimensional static
and spherically symmetric wormholes coupled with nonlinear
electrodynamics, and in Sec. \ref{RotWH} rotating traversable
wormholes in the context of nonlinear electrodynamics are studied.
In Sec. \ref{Conclusion} we conclude.

\section{Static and spherically symmetric
wormholes}\label{StaticWH}

\subsection{$(2+1)-$dimensional wormhole}

In this Section, we shall be interested in $(2+1)-$dimensional
general relativity coupled to nonlinear electrodynamics. We will
use geometrized units throughout this work, i.e., $G=c=1$. The
respective action is given by
\begin{equation}
S=\int \sqrt{-g}\left[\frac{R}{16\pi}+L(F)\right]\,d^3x  \,,
\end{equation}
where $R$ is the Ricci scalar and $L(F)$ is a gauge-invariant
electromagnetic Lagrangian, which we shall leave unspecified at
this stage, depending on the invariant $F$ given by
$F=\frac{1}{4}F^{\mu\nu}F_{\mu\nu}$.
$F_{\mu\nu}=A_{\nu,\mu}-A_{\mu,\nu}$ is the electromagnetic field.
Note that the factor $1/16\pi$, in the action, is maintained to
keep the parallelism with $(3+1)-$dimensional
theory~\cite{Garcia4}.

In Einstein-Maxwell theory, the Lagrangian is defined as
$L(F)\equiv -F/4\pi$, but here we consider more general choices of
electromagnetic Lagrangians, however, depending on the single
invariant $F$. It is perhaps important to emphasize that we do not
consider the case where $L$ depends on the invariant $G \equiv
\frac{1}{4}F_{\mu\nu}{}^*F^{\mu\nu}$, where $*$ denotes the Hodge
dual with respect to $g_{\mu\nu}$.

Varying the action with respect to the gravitational field
provides the Einstein tensor
\begin{equation}
G_{\mu\nu}=8\pi
(g_{\mu\nu}L-F_{\mu\alpha}F_{\nu}{}^{\alpha}\,L_{F}) \,,
\end{equation}
where $L_F\equiv d L/d F$. Clearly, the stress-energy tensor is
given by
\begin{equation}
T_{\mu\nu}=g_{\mu\nu}\,L(F)-F_{\mu\alpha}F_{\nu}{}^{\alpha}\,L_{F}\,,
    \label{stress-energy}
\end{equation}
where the Einstein field equation is defined as $G_{\mu\nu}=8\pi
T_{\mu\nu}$.
The variation of the action with respect to the electromagnetic
potential $A_\mu$, yields the electromagnetic field equations
\begin{eqnarray}
\left(F^{\mu\nu}\,L_{F}\right)_{;\mu}&=&0 \,,
     \label{em-field}
\end{eqnarray}
where the semi-colon denotes a covariant derivative.

The spacetime metric representing a spherically symmetric and
static $(2+1)-$dimensional wormhole is given by
\begin{equation}
ds^2=-e ^{2\Phi(r)}\,dt^2+\frac{dr^2}{1- b(r)/r}+r^2 \, d\phi ^2
\label{metricwormhole}\,,
\end{equation}
where $\Phi(r)$ and $b(r)$ are functions of the radial coordinate,
$r$. $\Phi(r)$ is denoted as the redshift function, for it is
related to the gravitational redshift; $b(r)$ is called the form
function \cite{Morris}. The radial coordinate has a range that
increases from a minimum value at $r_0$, corresponding to the
wormhole throat, to $\infty$.

For the wormhole to be traversable, one must demand the absence of
event horizons, which are identified as the surfaces with
$e^{2\Phi}\rightarrow 0$, so that $\Phi(r)$ must be finite
everywhere. A fundamental property of wormhole physics is the
flaring out condition, which is deduced from the mathematics of
embedding, and is given by $(b-b'r)/b^2>0$ \cite{Morris,Hochberg}.
Note that at the throat $b(r_0)=r=r_0$, the flaring out condition
reduces to $b'(r_0)<1$. The condition $(1-b/r)>0$ is also imposed.

Taking into account the symmetries of the geometry, we shall
consider the following electromagnetic tensor
\begin{equation}
F_{\mu\nu}=E(r)(\delta^t_\mu \delta^r_\nu-\delta^r_\mu
\delta^t_\nu)+B(r)(\delta^\phi_\mu \delta^r_\nu-\delta^r_\mu
\delta^\phi_{\nu})   \label{em-tensor}\,.
\end{equation}
Note that the only non-zero terms for the electromagnetic tensor
are the following $F_{tr}=-F_{rt}=E(r)$ and $F_{\phi
r}=-F_{r\phi}=B(r)$.
The invariant $F=F^{\mu\nu}F_{\mu\nu}/4$ is given by
\begin{eqnarray}
F=-\frac{1}{2}\;\left(1-\frac{b}{r}\right)\,\left[e^{-2\Phi}\,
E^2(r)-\frac{1}{r^2}\,B^2(r)\right]
     \,.
\end{eqnarray}
The electromagnetic field equation, Eq. (\ref{em-field}), provides
the following relationships
\begin{eqnarray}
e^{-\Phi}\,\left(1-\frac{b}{r}\right)^{1/2}E\,L_{F}&=&\frac{C_e}{r}
\,,
        \label{emf:electric}        \\
\frac{1}{r}\,\left(1-\frac{b}{r}\right)^{1/2}B\,L_{F}&=&C_m\,e^{-\Phi}
\,.    \label{emf:magnetic}
\end{eqnarray}
where the constants of integration $C_e$ and $C_m$ are related to
the electric and magnetic charge, $q_e$ and $q_m$, respectively.

The mathematical analysis and the physical interpretation will be
simplified using a set of orthonormal basis vectors. These may be
interpreted as the proper reference frame of a set of observers
who remain at rest in the coordinate system $(t,r,\phi)$, with
$(r,\phi)$ fixed.
Now, the non-zero components of the Einstein tensor,
$G_{\hat{\mu}\hat{\nu}}$, in the orthonormal reference frame, are
given by
\begin{eqnarray}
G_{\hat{t}\hat{t}}&=&\;\frac{b'r-b}{2r^3}   \label{Gtt}\,,\\
G_{\hat{r}\hat{r}}&=&\left(1-\frac{b}{r}\right) \frac{\Phi'}{r}  \label{Grr}\,,\\
G_{\hat{\phi}\hat{\phi}}&=& \left(1-\frac{b}{r}\right)\left[\Phi
''+ (\Phi')^2- \frac{b'r-b}{2r(r-b)}\Phi' \right] \label{Gpp}\,.
\end{eqnarray}
The Einstein field equation, $G_{\hat{\mu}\hat{\nu}}=8\pi
\,T_{\hat{\mu}\hat{\nu}}$, requires that the Einstein tensor be
proportional to the stress-energy tensor, so that in the
orthonormal basis the latter must have an identical algebraic
structure as the Einstein tensor components,
$G_{\hat{\mu}\hat{\nu}}$, i.e., Eqs. (\ref{Gtt})-(\ref{Gpp}).

Recall that a fundamental condition in wormhole physics is the
violation of the NEC, which is defined as
$T_{\mu\nu}k^{\mu}k^{\nu} \geq 0$, where $k^\mu$ is {\it any} null
vector. Considering the orthonormal reference frame with
$k^{\hat{\mu}}=(1,\pm 1,0)$, we have
\begin{equation}\label{NECthroat}
T_{\hat{\mu}\hat{\nu}}k^{\hat{\mu}}k^{\hat{\nu}}=
\frac{1}{8\pi}\,\left[\frac{b'r-b}{r^3}+
\left(1-\frac{b}{r}\right) \frac{\Phi '}{r} \right]  \,.
\end{equation}
Using the flaring out condition of the throat, $(b-b'r)/2b^2>0$
\cite{Morris,Visser}, and considering the finite character of
$\Phi(r)$, we verify that evaluated at the throat the NEC is
violated, i.e.,
$T_{\hat{\mu}\hat{\nu}}k^{\hat{\mu}}k^{\hat{\nu}}<0$. Matter that
violates the NEC is denoted as {\it exotic matter}.

The only non-zero components of $T_{\hat{\mu}\hat{\nu}}$, taking
into account Eq. (\ref{stress-energy}), are
\begin{eqnarray}
T_{\hat{t}\hat{t}}&=&-L-e^{-2\Phi}\left(1-\frac{b}{r}\right)\,E^2\,L_{F}
\,,
     \label{TttNLE}   \\
T_{\hat{r}\hat{r}}&=&L+e^{-2\Phi}\left(1-\frac{b}{r}\right)\,E^2\,L_{F}
            \nonumber   \\
&&-\left(1-\frac{b}{r}\right)\,\frac{B^2}{r^2}\,L_{F}    \,,
\label{TrrNLE}      \\
T_{\hat{\phi}\hat{\phi}}&=&L-\left(1-\frac{b}{r}\right)\,\frac{B^2}{r^2}\,L_{F}
\,. \label{TppNLE}
\end{eqnarray}
We need to impose the conditions
$|e^{-2\phi}(1-b/r)E^2L_{F}|<\infty$ and
$|(1-b/r)B^2L_{F}|<\infty$ as $r\rightarrow r_0$, to ensure the
regularity of the stress-energy tensor components.

Note that the Lagrangian may be obtained from the following
relationship:
$L=T_{\hat{\phi}\hat{\phi}}-T_{\hat{t}\hat{t}}-T_{\hat{r}\hat{r}}$,
and using the Einstein field equation, is given by
\begin{eqnarray}\label{Lag}
L&=&\frac{1}{8\pi} \Bigg\{\left(1-\frac{b}{r}\right)\Big[\Phi ''+
(\Phi')^2-\frac{\Phi'}{r}
      \nonumber      \\
&&- \frac{b'r-b}{2r(r-b)}\Phi' \Big] -\frac{b'r-b}{2r^3} \Bigg\}
\,.
\end{eqnarray}

However, from the metric (\ref{metricwormhole}) we verify the
following zero components of the Einstein tensor:
$G_{\hat{t}\hat{r}}=0$, $G_{\hat{r}\hat{\phi}}=0$ and
$G_{\hat{t}\hat{\phi}}=0$. Thus, through the Einstein field
equation, a further restriction may be obtained from
$T_{\hat{t}\hat{\phi}}=0$, i.e.,
\begin{equation}
T_{\hat{t}\hat{\phi}}=-\frac{1}{r}E(r)B(r)e^{-\phi}(1-b/r)\,L_F
\,,
\end{equation}
which imposes that $E(r)=0$ or $B(r)=0$, considering the
non-trivial case of $L_{F}$ non-zero. It is rather interesting
that both $E(r)$ and $B(r)$ cannot coexist simultaneously, in the
present $(2+1)-$dimensional case.

For the specific case of $B(r)=0$, from Eqs.
(\ref{Gtt})-(\ref{Grr}) and Eqs. (\ref{TttNLE})-(\ref{TrrNLE}), we
verify the following condition
\begin{equation}
\Phi'=-\frac{b'r-b}{2r(r-b)}\,,
\end{equation}
which may be integrated to yield the solution
\begin{equation}
e^{2\Phi}=\left(1-\frac{b}{r}\right)  \,.
\end{equation}
This corresponds to a non-traversable wormhole solution, as it
possesses an event horizon at the throat, $r=r_0$.

Now, consider the case of $E(r)=0$ and $B(r)\neq 0$. For this case
we have $T_{\hat{r}\hat{r}}=T_{\hat{\phi}\hat{\phi}}$, and the
respective Einstein tensor components, Eqs.
(\ref{Grr})-(\ref{Gpp}), provide the following differential
equation
\begin{equation}\label{diffeq}
\frac{\Phi'}{r}=\Phi ''+ (\Phi')^2- \frac{b'r-b}{2r(r-b)}\Phi' \,.
\end{equation}
Considering a specific choice of $b(r)$ or $\Phi(r)$, one may, in
principle, obtain a solution for the geometry. Equation
(\ref{diffeq}) may be formally integrated to yield the following
general solution
\begin{equation}\label{gen:Phi}
\Phi(r)=\ln\left[C_1 \int
r\left(1-\frac{b(r)}{r}\right)^{-1/2}\,dr + C_2 \right] \,,
\end{equation}
where $C_1$ and $C_2$ are constants of integration. For instance,
consider a constant form function, $b(r)=r_0$, so that from Eq.
(\ref{gen:Phi}), we deduce
\begin{eqnarray}\label{Phi:sol1}
\Phi(r)&=&\ln\Big\{C_2+\frac{C_1}{8}\Big[2\sqrt{r(r-r_0)}\,(2r+3r_0)
          \nonumber     \\
&&+3r_0^2\ln{\left(r-r_0/2+\sqrt{r(r-r_0)}\right)}\Big]\Big\} \,,
\end{eqnarray}
which at the throat reduces to
\begin{equation}
\Phi(r_0)=\ln\left[C_2+\frac{3C_1r_0^2}{8}\ln\left(\frac{r_0}{2}\right)\right]\,.
\end{equation}
To obtain a regular solution at the throat, we impose the
condition: $C_2+(3C_1r_0^2/8)\ln(r_0/2)>0$.

Consider for instance $b(r)=r_0^2/r$, then Eq. (\ref{gen:Phi})
provides the solution
\begin{eqnarray}\label{Phi:sol2}
\Phi(r)&=&\ln\Bigg\{C_2+\frac{C_1}{2}\Bigg[r\sqrt{r^2-r_0^2}
          \nonumber     \\
&&+r_0^2\ln{\left(r+\sqrt{r^2-r_0^2)}\right)}\Bigg]\Bigg\} \,,
\end{eqnarray}
which at the throat, reduces to
\begin{equation}
\Phi(r_0)=\ln\left[C_2+\frac{C_1r_0^2}{2}\ln(r_0)\right]\,.
\end{equation}
Once again, to ensure a regular solution, we need to impose the
following condition: $C_2+(C_1r_0^2/2)\ln(r_0)>0$. Note that these
specific solutions are not asymptotically flat, however, they may
be matched to an exterior vacuum spacetime, much in the spirit of
Refs. \cite{LLQ-PRD,wormhole-shell}.

However, a subtlety needs to be pointed out. Consider Eqs.
(\ref{Gtt})-(\ref{Grr}) and (\ref{TttNLE})-(\ref{TrrNLE}), from
which we deduce
\begin{equation}\label{dphi}
\Phi'=-\frac{b'r-b}{2r(r-b)}-\frac{8\pi B^2}{r} \,L_{F}\,.
\end{equation}
Now, taking into account Eq. (\ref{emf:magnetic}), we find the
following relationships for the magnetic field, $B(r)$, and for
$L_{F}$
\begin{equation}\label{magnetic-field}
B(r)=-\frac{e^{\Phi}}{8\pi
C_m}\left[\frac{b'r-b}{2r^2(1-b/r)^{1/2}}+\left(1-\frac{b}{r}\right)^{1/2}\Phi'
\right]  \,,
\end{equation}
and
\begin{equation}\label{magnetic-LF}
L_{F}=-\frac{8\pi C_m^2 r\,
e^{-2\Phi}}{\frac{b'r-b}{2r^2}+(1-b/r)\,\Phi'}  \,,
\end{equation}
respectively. Considering that the redshift $\Phi$ be finite
throughout the spacetime, one immediately verifies that the
magnetic field $B(r)$ is singular at the throat, which is
transparent considering the first term in square brackets in the
right hand side of Eq. (\ref{magnetic-field}). This is an
extremely troublesome aspect of the geometry, as in order to
construct a traversable wormhole, singularities appear in the
physical fields. This aspect is in contradiction to the model
construction of nonlinear electrodynamics, founded on a principle
of finiteness, that a satisfactory theory should avoid physical
quantities becoming infinite \cite{BI}. Thus, one should impose
that these physical quantities be non-singular, and in doing so,
we verify that the general solution corresponds to a
non-traversable wormhole geometry. This may be verified by
integrating Eq. (\ref{dphi}), which yields the following general
solution
\begin{equation}
e^{2\Phi}=\left(1-\frac{b}{r} \right)\exp \left(-16\pi\int
\frac{B^2}{r}\,L_{F} \,dr \right) \,.
\end{equation}
We have considered the factor $|B^2L_{F}|<\infty$ as $r\rightarrow
r_0$, to ensure the regularity of the term in the exponential.
However, this solution corresponds to a non-traversable wormhole
solution, as it possesses an horizon at the throat, $b=r=r_0$.

One may also prove the non-existence of $(2+1)-$dimensional static
and spherically symmetric traversable wormholes in nonlinear
electrodynamics, through an analysis of the NEC violation. In the
context of nonlinear electrodynamics, and taking into account Eqs.
(\ref{TttNLE})-(\ref{TrrNLE}), we verify
\begin{equation}
T_{\hat{\mu}\hat{\nu}}k^{\hat{\mu}}k^{\hat{\nu}}
=-\left(1-\frac{b}{r}\right)\,\frac{B^2}{r^2}\,L_{F}
 \,,
\end{equation}
which evaluated at the throat, considering the regularity of $B$
and $L_{F}$, is identically zero, i.e.,
$T_{\hat{\mu}\hat{\nu}}k^{\hat{\mu}}k^{\hat{\nu}}|_{r_0} =0$. The
NEC is not violated at the throat, so that the flaring-out
condition is not satisfied, showing, therefore, the non-existence
of $(2+1)-$dimensional static and spherically symmetric
traversable wormholes in nonlinear electrodynamics.

\subsection{$(3+1)-$dimensional wormhole}

The action of $(3+1)-$dimensional general relativity coupled to
nonlinear electrodynamics is given by
\begin{equation}
S=\int \sqrt{-g}\left[\frac{R}{16\pi}+L(F)\right]\,d^4x  \,,
\end{equation}
where $R$ is the Ricci scalar and the gauge-invariant
electromagnetic Lagrangian, $L(F)$, depends on a single invariant
$F$ \cite{Pleb,Pleb2}, defined by $F\equiv
\frac{1}{4}F^{\mu\nu}F_{\mu\nu}$, as before. We shall not consider
the case where $L$ depends on the invariant $G \equiv
\frac{1}{4}F_{\mu\nu}{}^*F^{\mu\nu}$, as mentioned in the
$(2+1)-$dimensional case.

Varying the action with respect to the gravitational field
provides the Einstein field equations $G_{\mu\nu}=8\pi
T_{\mu\nu}$, where the stress-energy tensor is given by
\begin{equation}
T_{\mu\nu}=g_{\mu\nu}\,L(F)-F_{\mu\alpha}F_{\nu}{}^{\alpha}\,L_{F}\,.
    \label{4dim-stress-energy}
\end{equation}
Taking into account the symmetries of the geometry, the only
non-zero compatible terms for the electromagnetic tensor are
$F_{tr}=E(x^\mu)$ and $F_{\theta\phi}=B(x^\mu)$.

The spacetime metric representing a spherically symmetric and
static $(3+1)-$dimensional wormhole takes the form \cite{Morris}
\begin{equation}
ds^2=-e ^{2\Phi(r)}\,dt^2+\frac{dr^2}{1- b(r)/r}+r^2 \,(d\theta
^2+\sin ^2{\theta} \, d\phi ^2) \label{4metricwormhole}\,.
\end{equation}
The non-zero components of the Einstein tensor, given in an
orthonormal reference frame, are given by
\begin{eqnarray}
G_{\hat{t}\hat{t}}&=& \,\frac{b'}{r^2} \label{rhoWH} \,, \\
G_{\hat{r}\hat{r}}&=& -\frac{b}{r^3}+2 \left(1-\frac{b}{r}
\right) \frac{\Phi'}{r}  \label{prWH} \,, \\
G_{\hat{\phi}\hat{\phi}}&=&G_{\hat{\theta}\hat{\theta}}=
\left(1-\frac{b}{r}\right)\Bigg[\Phi ''+ (\Phi')^2-
\frac{b'r-b}{2r(r-b)}\Phi'
    \nonumber    \\
&&\hspace{1.2cm}-\frac{b'r-b}{2r^2(r-b)}+\frac{\Phi'}{r} \Bigg]
\label{ptWH}\,.
\end{eqnarray}
Its a simple matter to prove that for this geometry, the NEC is
identical to Eq. (\ref{NECthroat}), and is also violated at the
throat, i.e.,
$T_{\hat{\mu}\hat{\nu}}k^{\hat{\mu}}k^{\hat{\nu}}<0$.

The relevant components for the stress-energy tensor, regarding
the analysis of the NEC, are the following
\begin{eqnarray}
T_{\hat{t}\hat{t}}&=&-L-e^{-2\Phi}\left(1-\frac{b}{r}\right)\,E^2\,L_{F}
\,,
     \label{4TttNLE}   \\
T_{\hat{r}\hat{r}}&=&L+e^{-2\Phi}\left(1-\frac{b}{r}\right)\,E^2\,L_{F}
              \,.
\label{4TrrNLE}
\end{eqnarray}
Analogously with the $(2+1)-$dimensional case, we will consider
$|e^{-2\Phi}(1-b/r)\,E^2\,L_{F}|<\infty$, as $r\rightarrow r_0$,
to ensure that the stress-energy tensor components are regular.

From Eqs. (\ref{rhoWH})-(\ref{prWH}) and Eqs.
(\ref{4TttNLE})-(\ref{4TrrNLE}), we verify the following condition
\begin{equation}
\Phi'=-\frac{b'r-b}{2r(r-b)}\,,
\end{equation}
which may be integrated to yield the solution $e^{2\Phi}=(1-b/r)$,
rendering a non-traversable wormhole solution, as it possesses an
event horizon at the throat, $r=r_0$.

Note that the NEC, for the stress-energy tensor defined by
(\ref{4dim-stress-energy}), is identically zero for arbitrary $r$,
i.e., $T_{\hat{\mu}\hat{\nu}}k^{\hat{\mu}}k^{\hat{\nu}}=0$. In
particular this implies that the flaring-out condition of the
throat is not satisfied, showing, therefore, the non-existence of
$(3+1)-$dimensional static and spherically symmetric traversable
wormholes coupled to nonlinear electrodynamics. The analysis
outlined in this Section is consistent with that of Refs.
\cite{Bronnikov1,Bronnikov2}, where it was pointed out that
nonlinear electrodynamics, with any Lagrangian of the form $L(F)$,
coupled to general relativity cannot support static and
spherically symmetric $(3+1)-$dimensional traversable wormholes.

The impediment to the construction of traversable wormholes may be
overcome by considering a non-interacting anisotropic distribution
of matter coupled to nonlinear electrodynamics. This may be
reflected by the following superposition of the stress-energy
tensor
\begin{equation}
T_{\mu\nu}=T_{\mu\nu}^{\rm fluid}+T_{\mu\nu}^{\rm NED} \,,
\end{equation}
where $T_{\mu\nu}^{\rm NED}$ is given by Eq.
(\ref{4dim-stress-energy}), and $T_{\mu\nu}^{\rm fluid}$ is
provided by
\begin{equation}
T_{\mu\nu}^{\rm fluid}=(\rho+p_t)U_\mu \, U_\nu+p_t\,
g_{\mu\nu}+(p_r-p_t)\chi_\mu \chi_\nu \,.
\end{equation}
$U^\mu$ is the four-velocity and $\chi^\mu$ is the unit spacelike
vector in the radial direction. $\rho(r)$ is the energy density,
$p_r(r)$ is the radial pressure measured in the direction of
$\chi^\mu$, and $p_t(r)$ is the transverse pressure measured in
the orthogonal direction to $\chi^\mu$.

Now, the NEC takes the form
\begin{eqnarray}
T_{\hat{\mu}\hat{\nu}}k^{\hat{\mu}}k^{\hat{\nu}}&=&\rho(r)+p_r(r)
          \nonumber   \\
&=&\frac{1}{8\pi}\,\left[\frac{b'r-b}{r^3}+
\left(1-\frac{b}{r}\right) \frac{\Phi '}{r} \right]  \,,
\end{eqnarray}
which evaluated at the throat, reduces to the NEC violation
analysis of Ref. \cite{Morris}, i.e., $\rho+p_r<0$.

\section{Stationary and axisymmetric wormholes}\label{RotWH}

\subsection{$(2+1)-$dimensional wormhole}

We now analyze nonlinear electrodynamics coupled to a stationary
axisymmetric $(2+1)-$dimensional wormhole geometry. The stationary
character of the spacetime implies the presence of a time-like
Killing vector field, generating invariant time translations. The
axially symmetric character of the geometry implies the existence
of a spacelike Killing vector field, generating invariant
rotations with respect to the angular coordinate $\phi$. Consider
the metric
\begin{equation}\label{rwhmn}
  ds^2=-N^2dt^2+\frac{dr^2}{1-b/r}+r^2K^2(d\phi-\omega\,dt)^2  \,,
\end{equation}
where $N, K, \omega$ and $b$ are functions of $r$. $\omega(r)$ may
be interpreted as the angular velocity $ d\phi/ dt$ of a particle.
$N$ is the analog of the redshift function in Eq.
(\ref{metricwormhole}) and is finite and nonzero to ensure that
there are no event horizons. We shall also assume that $K(r)$ is a
positive, nondecreasing function of $r$ that determines the proper
radial distance $R$, i.e.,
\begin{equation}
R\equiv rK\,,\qquad R'>0\,.
\end{equation}

To transform to an orthonormal reference frame, the one-forms in
the orthonormal basis transform as
$\tilde{\Theta}^{\hat{\mu}}=\Lambda^{\hat{\mu}}{}_{\nu}\,\Theta^\nu$.
The metric (\ref{rwhmn}) can be diagonalized
\begin{equation}\label{rwhmo}
  ds^2=-(\Theta^{\hat{t}})^2+(\Theta^{\hat{r}})^2+(\Theta^{\hat{\phi}})^2
  \,,
\end{equation}
by means of the tetrad
\begin{eqnarray}\label{tet2}
  \Theta^{\hat{t}}&=&Ndt  \,,  \\
  \Theta^{\hat{r}}&=&(1-b/r)^{-1/2}dr \,,  \\
  \Theta^{\hat{\phi}}&=&rK(d\phi-\omega dt)  \,.
\end{eqnarray}
Now, $\Lambda^{\mu}{}_{\hat{\alpha}} \;
\Lambda^{\hat{\alpha}}{}_{\nu} = \delta^{\mu}{}_{\nu}$ and
$\Lambda^{\mu}{}_{\hat{\nu}}$ is defined as
\begin{equation}
(\Lambda^{\mu}{}_{\hat{\nu}})=\left[
\begin{array}{ccc}
1/N&0&0 \\
0&(1-b/r)^{1/2}&0 \\
\omega/N&0&(rK)^{-1}
\end{array}
\right]   \label{tranfs3}\,.
\end{equation}
From the latter transformation, one may deduce the orthonormal
basis vectors, ${\bf
e}_{\hat{\mu}}=\Lambda^{\nu}{}_{\hat{\mu}}\,{\bf e}_{\nu}$, given
by
\begin{eqnarray}
{\bf e}_{\hat{t}}&=&\frac{1}{N}\,{\bf e}_{t}+\frac{\omega}{N}\,{\bf e}_{\phi} \,,  \\
{\bf e}_{\hat{r}}&=&\left(1-\frac{b}{r}\right)^{1/2}\,{\bf e}_{r}  \,, \\
{\bf e}_{\hat{\phi}}&=&\frac{1}{rK}\,{\bf e}_{\phi} \,.
\end{eqnarray}
Using the fact that ${\bf e}_{\alpha}\cdot{\bf
e}_{\beta}=g_{\alpha\beta}$, we have ${\bf e}_{\hat{\mu}}\cdot{\bf
e}_{\hat{\nu}}=g_{\hat{\mu}\hat{\nu}}=\eta_{\hat{\mu}\hat{\nu}}$.

Using the Einstein field equation,
$G_{\hat\mu\hat\nu}=8\pi\,T_{\hat\mu\hat\nu}$ and taking into
account the null vector $k^{\hat\mu}\,=(1,\pm 1,0)$, we obtain the
following relationship, at the throat
\begin{equation}\label{gnec}
  T_{\hat\mu\hat\nu}k^{\hat\mu}\,k^{\hat\nu} =-\frac{1-b'}{16\pi
  r_0^2K}(K+r_0K')\,.
\end{equation}
We verify that the NEC is clearly violated because the conditions
$K>0$ and $K'>0$ are imposed by construction \cite{Teo}, so that
the metric (\ref{rwhmn}) can describe a wormhole type solution.

\bigskip
Now for nonlinear electrodynamics, we consider the stress energy
tensor given by Eq. (\ref{stress-energy}), where the nonzero
components of the electromagnetic tensor are
\begin{equation}\label{2emtcomp}
  F_{tr}=-F_{rt}, \quad F_{t\phi}=-F_{\phi t}, \quad F_{\phi
  r}=-F_{r\phi}\,,
\end{equation}
which are only functions of the radial coordinate $r$. Then, using
the orthonormal reference frame, the NEC takes the following form
\begin{equation}\label{2setnec}
  T_{\hat\mu\hat\nu}k^{\hat\mu}\,k^{\hat\nu} =
  -\left[N^2\left(1-\frac{b}{r}\right)F_{tr}^2+F_{t\phi}^2\right]\frac{L_{F}}{r^2K^2N^2}
  \,,
\end{equation}
which, at the throat, reduces to
\begin{equation}\label{2setnect}
 T_{\hat\mu\hat\nu}k^{\hat\mu}\,k^{\hat\nu}\big|_{r_0} =
 -\frac{1}{r_0^2K^2N^2}F_{t\phi}^2\,L_{F}  \,.
\end{equation}
From this relationship, we verify that the NEC is violated at the
throat only if the derivative $L_{F}$ is positive and $F_{t\phi}$
is nonzero. The latter condition, $F_{t\phi}\neq 0$, is imposed to
have a compatibility of Eqs. (\ref{gnec}) and (\ref{2setnect}).

However, note that from the metric (\ref{rwhmn}) we verify the
following zero components of the Einstein tensor: $G_{tr}=0$ and
$G_{r\phi}=0$, implying that $T_{tr}=0$ and $T_{r\phi}=0$. These
stress-energy tensor components are given by
\begin{eqnarray}
T_{tr}&=&-(F_{tr}\,g^{t \phi}+F_{r \phi}\, g^{\phi\phi})\,F_{t
\phi}\,L_{F}\,,    \\
T_{r\phi}&=&-(F_{tr}\, g^{tt}+F_{r \phi}\, g^{t\phi})\,F_{t
\phi}\,L_{F}\,.
\end{eqnarray}
From these conditions, considering that the derivative $L_{F}$ be
finite and positive and the non-trivial case $F_{t\phi}\neq 0$, we
find that
\begin{equation}
(g^{t\phi})^2=g^{tt}g^{\phi\phi}  \,.
\end{equation}
From the above imposition we deduce $N=0$, implying the presence
of an event horizon, showing the non-existence of
$(2+1)-$dimensional stationary and axially symmetric traversable
wormholes coupled to nonlinear electrodynamics.

\subsection{$(3+1)-$dimensional wormhole}

Now, consider the stationary and axially symmetric
$(3+1)-$dimensional spacetime, and analogously to the previous
case, it possesses a time-like Killing vector field, which
generates invariant time translations, and a spacelike Killing
vector field, which generates invariant rotations with respect to
the angular coordinate $\phi$. We have the following metric
\begin{equation}\label{3rwh}
  ds^2=-N^2dt^2+e^{\mu}\,dr^2+r^2K^2[d\theta^2+\sin^2\theta(d\phi-\omega\,dt)^2]
\end{equation}
where $N$, $K$, $\omega$ and $\mu$ are functions of $r$ and
$\theta$~\cite{Teo}. $\omega(r,\theta)$ may be interpreted as the
angular velocity $ d\phi/ dt$ of a particle that falls freely from
infinity to the point $(r,\theta)$.
For simplicity, we shall consider the definition \cite{Teo}
\begin{equation}
e^{-\mu(r,\theta)}=1-\frac{b(r,\theta)}{r}\,,
\end{equation}
which is well suited to describe a traversable wormhole.
Assume that $K(r,\theta)$ is a positive, nondecreasing function of
$r$ that determines the proper radial distance $R$, i.e., $R\equiv
rK$ and $R_r>0$ \cite{Teo}, as for the $(2+1)-$dimensional case.
We shall adopt the notation that the subscripts $_r$ and
$_{\theta}$ denote the derivatives in order of $r$ and ${\theta}$,
respectively \cite{Teo}.

We shall also write down the contravariant metric tensors, which
will be used later, and are given by
\begin{eqnarray}\label{contravar}
&&g^{tt}=-\frac{1}{N^2}\,, \quad
g^{rr}=\left(1-\frac{b}{r}\right)\,, \quad
g^{\theta\theta}=\frac{1}{r^2K^2}\,,
      \nonumber     \\
&&g^{\phi\phi}=\frac{N^2-r^2\omega^2K^2\sin^2\theta}{r^2N^2K^2\sin^2\theta}
\,,    \quad     g^{t\phi}=-\frac{\omega}{N^2}\,.
\end{eqnarray}

Note that an event horizon appears whenever $N=0$~\cite{Teo}. The
regularity of the functions $N$, $b$ and $K$ are imposed, which
implies that their $\theta$ derivatives vanish on the rotation
axis, $\theta=0,\,\pi$, to ensure a non-singular behavior of the
metric on the rotation axis. The metric (\ref{3rwh}) reduces to
the Morris-Thorne spacetime metric (\ref{metricwormhole}) in the
limit of zero rotation and spherical symmetry
\begin{eqnarray}
&N(r,\theta)\rightarrow{\rm e}^{\Phi(r)},\quad
b(r,\theta)\rightarrow b(r)\,,
       \\
&K(r,\theta)\rightarrow1\,, \quad \omega(r,\theta)\rightarrow0\,.
\end{eqnarray}
In analogy with the Morris-Thorne case, $b(r_0)=r_0$ is identified
as the wormhole throat, and the factors $N$, $K$ and $\omega$ are
assumed to be well-behaved at the throat.

The scalar curvature of the space-time (\ref{3rwh}) is extremely
messy, but at the throat $r=r_0$ simplifies to
\begin{eqnarray}\label{rotWHRicciscalar}
R&=&-\frac{1}{r^2K^2}\left(\mu_{\theta\theta}
+\frac{1}{2}\mu_\theta^2\right)
-\frac{\mu_\theta}{Nr^2K^2}\,\frac{(N
\sin\theta)_\theta}{\sin\theta}
           \nonumber  \\
&&-\frac{2}{Nr^2K^2}\,\frac{(N_{\theta}
\sin\theta)_\theta}{\sin\theta}
-\frac{2}{r^2K^3}\,\frac{(K_\theta \sin\theta)_\theta}{\sin\theta}
           \nonumber    \\
&&+e^{-\mu}\,\mu_r\,\left[\ln(Nr^2K^2)\right]_r
+\frac{\sin^2\theta\,\omega_\theta^2}{2N^2}
           \nonumber  \\
&&+\frac{2}{r^2K^4}\,(K^2+K_\theta^2)  \,.
\end{eqnarray}
The only troublesome terms are the ones involving the terms with
$\mu_\theta$ and $\mu_{\theta\theta}$, i.e.,
\begin{equation}
\mu_\theta=\frac{b_\theta}{(r-b)}\,,   \qquad \mu_{\theta\theta}
+\frac{1}{2}\mu_\theta^2=\frac{b_{\theta\theta}}{r-b}
+\frac{3}{2}{b_\theta{}^2\over(r-b)^2}\,.
\end{equation}
Note that one needs to impose that $b_\theta=0$ and
$b_{\theta\theta}=0$ at the throat to avoid curvature
singularities. This condition shows that the throat is located at
a constant value of $r$.

Thus, one may conclude that the metric (\ref{3rwh}) describes a
rotating wormhole geometry, with an angular velocity $\omega$. The
factor $K$ determines the proper radial distance. $N$ is the
analog of the redshift function in Eq. (\ref{4metricwormhole}) and
is finite and nonzero to ensure that there are no event horizons
or curvature singularities. $b$ is the shape function which
satisfies $b\leq r$; it is independent of $\theta$ at the throat,
i.e., $b_\theta=0$; and obeys the flaring out condition $b_r<1$.

\medskip

In the context of nonlinear electrodynamics, we consider the
stress energy tensor defined in Eq. (\ref{4dim-stress-energy}).
The nonzero components of the electromagnetic tensor are
\begin{eqnarray}\label{3emtcomp}
 && F_{tr}=-F_{rt},F_{t\theta}=-F_{\theta t},F_{t\phi}=-F_{\phi t},
  \\
 && F_{\phi r}=-F_{r\phi},F_{r\theta}=-F_{\theta r},F_{\theta\phi}=-F_{\phi\theta}
\end{eqnarray}
which are functions of the radial coordinate $r$ and the angular
coordinate $\theta$.

The analysis is simplified using an orthonormal reference frame,
with the following orthonormal basis vectors
\begin{eqnarray}
{\bf e}_{\hat{t}}&=&\frac{1}{N}\,{\bf e}_{t}+\frac{\omega}{N}\,{\bf e}_{\phi} \,,  \\
{\bf e}_{\hat{r}}&=&\left(1-\frac{b}{r}\right)^{1/2}\,{\bf e}_{r}  \,, \\
{\bf e}_{\hat{\theta}}&=&\frac{1}{rK}\,{\bf e}_{\theta}  \,,  \\
{\bf e}_{\hat{\phi}}&=&\frac{1}{rK\sin\theta}\,{\bf e}_{\phi} \,.
\end{eqnarray}
Now the Einstein tensor components are extremely messy, but assume
a more simplified form using the orthonormal reference frame and
evaluated at the throat. They have the following non-zero
components
\begin{eqnarray}
G_{\hat{t}\hat{t}}&=&-\frac{(K_\theta
\sin\theta)_\theta}{r^2K^3\sin\theta}
-\frac{\omega_\theta^2\,\sin^2\theta}{4N^2}
+e^{-\mu}\,\mu_r\,\frac{(rK)_r}{rK}
       \nonumber     \\
&&+\frac{K^2+K_\theta^2}{r^2K^4}
    \,,   \label{rotGtt}
\\
G_{\hat{r}\hat{r}}&=&\frac{(K_\theta
\sin\theta)_\theta}{r^2K^3\sin\theta}
-\frac{\omega_\theta^2\,\sin^2\theta}{4N^2}
+\frac{(N_\theta \sin\theta)_\theta}{Nr^2K^2\sin\theta}
         \nonumber    \\
&&-\frac{K^2+K_\theta^2}{r^2K^4}
   \,,
\\
G_{\hat{r}\hat{\theta}}&=&\frac{e^{-\mu/2}\,\mu_\theta\,(rKN)_r}{2Nr^2K^2}
\,,
\\
G_{\hat{\theta}\hat{\theta}}&=& \frac{N_\theta(K
\sin\theta)_\theta}{Nr^2K^3\sin\theta}
+\frac{\omega_\theta^2\,\sin^2\theta}{4N^2}
     \\  \nonumber
&& -\frac{\mu_r\,e^{-\mu}(NrK)_r}{2NrK}
   \,,
\\
G_{\hat{\phi}\hat{\phi}}&=&
-\frac{\mu_r\,e^{-\mu}\,(NKr)_r}{2NKr}-\frac{3\sin^2\theta\,\omega_\theta^2}{4N^2}
         \nonumber   \\
&&+\frac{N_{\theta\theta}}{Nr^2K^2}-\frac{N_{\theta}K_{\theta}}{Nr^2K^3}
\,,    \label{rotGphiphi}
\\
G_{\hat{t}\hat{\phi}}&=&\frac{1}{4N^2K^2r}\;\Big(6NK\,\omega_{\theta}\,\cos\theta
+2NK\,\sin \theta\,\omega_{\theta \theta}
       \nonumber     \\
&&-\mu_{r}e^{-\mu}r^2NK^3\,\sin\theta\; \omega_{r}
+4N\,\omega_\theta\,\sin\theta\,K_\theta
      \nonumber   \\
&& -2K\,\sin\theta\,N_{\theta}\,\omega_{\theta}  \Big)   \,.
           \label{rotGtphi}
\end{eqnarray}
Note that the component $G_{\hat{r}\hat{\theta}}$ is zero at the
throat, however, we have included this term, as it shall be
helpful in the analysis of the stress-energy tensor components,
outlined below.

Using the Einstein field equation, the components
$T_{\hat{t}\hat{t}}$ and $T_{\hat{i}\hat{j}}$ have the usual
physical interpretations, and in particular,
$T_{\hat{t}\hat{\phi}}$ characterizes the rotation of the matter
distribution. It is interesting to note that constraints on the
geometry, placing restrictions on the stress energy tensor needed
to generate a general stationary and axisymmetric spacetime, were
found in Ref. \cite{Berg}. Taking into account the Einstein tensor
components above, the NEC at the throat is given by
\begin{eqnarray}\label{NEC}
8\pi\,T_{\hat{\mu} \hat{\nu}}k^{\hat{\mu}} k^{\hat{\nu}}&=&{\rm
e}^{-\mu}\mu_r{(rK)_r\over rK}
-{\omega_\theta{}^2\sin^2\theta\over2N^2}
      \nonumber     \\
&&+{(N_\theta\sin\theta)_\theta\over(rK)^2N\sin\theta}\,.
\end{eqnarray}
Rather than reproduce the analysis here, we refer the reader to
Ref. \cite{Teo}, where it was shown that the NEC is violated in
certain regions, and is satisfied in others. Thus, it is possible
for an infalling observer to move around the throat, and avoid the
exotic matter supporting the wormhole. However, it is important to
emphasize that one cannot avoid the use of exotic matter
altogether.

Using the stress-energy tensor, Eq. (\ref{4dim-stress-energy}), we
verify the following relationship
\begin{eqnarray}\label{3setnec}
T_{\hat\mu\hat\nu}k^{\hat\mu}k^{\hat\nu} &=&
-\Big[F_{t\phi}^2+\sin^2\theta(F_{t\theta}+\omega
F_{\phi\theta})^2
          \nonumber    \\
&&\hspace{-2.0cm}+\left(1-\frac{b}{r}\right)N^2(F_{\phi
r}^2+\sin^2\theta
F_{r\theta}^2)\Big]\frac{L_{F}}{r^2K^2N^2\sin^2\theta}   \,,
\end{eqnarray}
which evaluated at the throat reduces to
\begin{equation}\label{3setnect}
 T_{\hat\mu\hat\nu}k^{\hat\mu}k^{\hat\nu}= -[F_{t\phi}^2
 +\sin^2\theta (F_{t\theta}+\omega F_{\phi\theta})^2]
 \frac{L_{F}}{r_0^2K^2N^2\sin^2\theta}  \,.
\end{equation}
Note that for this expression to be compatible with Eq.
(\ref{NEC}), $L_{F}$ may be either positive, negative or zero.

\bigskip

The non-zero components of the Einstein tensor are precisely the
components expressed in Eqs. (\ref{rotGtt})-(\ref{rotGtphi}), so
that through the Einstein field equation, we have the following
zero components for the stress energy tensor in the
$(3+1)-$dimensional case: $T_{tr}=T_{t\theta}=T_{\phi
r}=T_{\phi\theta}=0$. Thus, taking into account this fact, and
considering that $L_{F}$ is regular, we obtain the following
relationships
\begin{eqnarray}
g^{\theta\theta}F_{t\theta}F_{r\theta}&=&F_{t\phi}(g^{t\phi}F_{tr}+g^{\phi\phi}F_{\phi
r})   \,,    \label{1}
        \\   \label{2}
-g^{\theta\theta}F_{\phi\theta}F_{r\theta}&=&F_{t\phi}(g^{tt}F_{tr}+g^{t\phi}F_{\phi
r})    \,,        \\   \label{3}
-g^{rr}F_{tr}F_{r\theta}&=&F_{t\phi}(g^{t\phi}F_{t\theta}+g^{\phi\phi}F_{\phi\theta})
\,,
        \\    \label{4}
g^{rr}F_{\phi
r}F_{r\theta}&=&F_{t\phi}(g^{tt}F_{t\theta}+g^{t\phi}F_{\phi\theta})
\,.
\end{eqnarray}
Now, rewriting Eqs. (\ref{1})-(\ref{2}) in order of $F_{t\theta}$
and $F_{\phi\theta}$, respectively, and introducing these in Eq.
(\ref{3}), we finally arrive at the following relationship
\begin{equation}
-g^{rr}N^2\sin^2\theta\,F_{r\theta}^2=F_{t\phi}^2 \,,
\end{equation}
from which we obtain that $(1-b/r)<0$, implying that $r<b$ for all
values of $r$ except at the throat. However, this restriction is
in clear contradiction with the definition of a traversable
wormhole, where the condition $(1-b/r)>0$ is imposed.

The same restriction can be inferred from the electromagnetic
field equations $(F^{\mu\nu}L_{F})_{;\mu}=0$, which provide the
following relationships
\begin{eqnarray}\label{3eqLFt}
&&  F^{rt}[\ln(L_{F})]_{,r}+F^{\theta
t}[\ln(L_{F})]_{,\theta}=-F^{\mu t}{}_{;\mu}   \,,
      \\
\label{3eqLFphi} &&
F^{r\phi}[\ln(L_{F})]_{,r}+F^{\theta\phi}[\ln(L_{F})]_{,\theta}
=-F^{\mu\phi}{}_{;\mu}    \,.
\end{eqnarray}
These can be rewritten in the following manner
\begin{eqnarray}
&&[\ln(L_{F})]_{,r}=\frac{F^{\theta
t}F^{\mu\phi}{}_{;\mu}-F^{\theta \phi}F^{\mu
t}{}_{;\mu}}{F^{rt}F^{\theta\phi}-F^{r \phi}F^{\theta t}} \,,
     \\
&&[\ln(L_{F})]_{,\theta}=\frac{F^{r\phi}F^{\mu t
}{}_{;\mu}-F^{rt}F^{\mu \phi}{}_{;\mu}}{F^{rt}F^{\theta\phi}-F^{r
\phi}F^{\theta t}} \,.
\end{eqnarray}
Now, a crucial point to note is that to have a solution, the term
in the denominator, $F^{rt}F^{\theta\phi}-F^{r\phi}F^{\theta t}$,
should be non-zero, and can be expressed as
\begin{eqnarray}\label{1p}
F^{rt}F^{\theta\phi}-F^{r\phi}F^{\theta
t}&=&g^{rr}g^{\theta\theta}\left[g^{tt}g^{\phi\phi}-(g^{t\phi})^2\right]
\times
         \nonumber \\
&&\times (F_{tr}F_{\phi\theta}-F_{t\theta}F_{\phi r })   \,.
\end{eqnarray}
However, using the Eqs. (\ref{1})-(\ref{4}), we may obtain an
alternative relationship, given by
\begin{equation}\label{2p}
  F^{rt}F^{\theta\phi}-F^{r\phi}F^{\theta
t}=\left(g^{rr}g^{\theta\theta}\frac{F_{r\theta}}{F_{t\phi}}\right)^2
(F_{tr}F_{\phi\theta}-F_{t\theta}F_{\phi r})   \,.
\end{equation}
Confronting both relationships, we obtain that $(1-b/r)<0$,
implying that $r<b$ for all values of $r$ except at the throat,
which, as before, is in clear contradiction with the definition of
a traversable wormhole.

If we consider the individual cases of $F_{tr}=0$, $F_{\phi r}=0$,
$F_{t \theta}=0$ or $F_{\phi \theta}=0$, separately, it is a
simple matter to verify that Eqs. (\ref{1})-(\ref{4}) impose the
restriction $N^2<0$, which does not satisfy the wormhole
conditions.

For the specific case of $F_{t \phi}=0$ (with $F_{r\theta} \neq
0$), the restrictions $F_{tr}=F_{\phi r}=F_{t \theta}=F_{\phi
\theta}=0$ are imposed. Taking into account these impositions one
verifies that from Eq. (\ref{3setnect}), we have
$T_{\hat\mu\hat\nu}k^{\hat\mu}k^{\hat\nu}=0$, which is not
compatible with the geometric conditions imposed by Eq.
(\ref{NEC}).

Considering $F_{r \theta}=0$ (with $F_{t\phi} \neq 0$), one
readily verifies from Eqs. (\ref{1})-(\ref{4}) the existence of an
event horizon, i.e., $N=0$, rendering the wormhole geometry
non-traversable.

The specific case of $F_{t\phi}=0$ and $F_{r\theta}=0$ also obeys
Eqs. (\ref{1})-(\ref{4}), and needs to be analyzed separately. To
show that this case is also in contradiction to the wormhole
conditions at the throat, we shall consider the following
stress-energy tensor components, for $F_{t\phi}=0$ and
$F_{r\theta}=0$
\begin{eqnarray}
T_{\hat{t}\hat{t}}&=&-L-\frac{(1-b/r)}{N^2}(F_{tr}+\omega F_{\phi
r })^2L_F
    \\  \nonumber
&&-\frac{1}{N^2r^2K^2}(F_{t\theta}+\omega F_{\phi \theta})^2L_F
\,,
  \label{SET1}  \\
T_{\hat{r}\hat{r}}&=&L+\frac{(1-b/r)}{N^2}(F_{tr}+\omega F_{\phi r
})^2L_F
    \\  \nonumber
&&-\frac{(1-b/r)}{r^2K^2\sin^2\theta}F_{\phi r}^2\,L_F     \,,
   \label{SET2} \\
T_{\hat{\theta}\hat{\theta}}&=&L+\frac{1}{N^2r^2K^2}(F_{t\theta}+\omega
F_{\phi \theta})^2L_F
    \\  \nonumber
&&-\frac{1}{r^4K^4\sin^2\theta}F_{\phi\theta}^2\;L_F   \,,
   \label{SET3} \\
T_{\hat{\phi}\hat{\phi}}&=&L-\frac{(1-b/r)}{r^2K^2\sin^2\theta}F_{\phi
r }^2\;L_F
      \\  \nonumber
&&-\frac{1}{r^4K^4\sin^2\theta}F_{\phi\theta}^2\;L_F  \,.
    \label{SET4}
\end{eqnarray}
An important result deduced from the above components is the
following
\begin{equation}\label{SETcond}
T_{\hat{t}\hat{t}}+T_{\hat{r}\hat{r}}+T_{\hat{\theta}\hat{\theta}}-
T_{\hat{\phi}\hat{\phi}}=0
\end{equation}
for all points in the geometry.

Now, considering the Einstein tensor components, Eqs.
(\ref{rotGtt})-(\ref{rotGphiphi}), evaluated at the throat, along
the rotation axis, and using the Einstein field equation, we
verify the following relationship
\begin{equation}\label{Einstein-cond}
G_{\hat{t}\hat{t}}+G_{\hat{r}\hat{r}}+G_{\hat{\theta}\hat{\theta}}-
G_{\hat{\phi}\hat{\phi}}=\frac{e^{-\mu}\mu_r(rK)_r}{rK}  \,,
\end{equation}
which is always positive. We have taken into account that the
functions $N$ and $K$ are regular, so that their $\theta$
derivatives vanish along the rotation axis, $\theta=0,\;\pi$, as
emphasized above. Therefore, through the Einstein field equation,
we verify that relationship (\ref{Einstein-cond}) is not
compatible with condition (\ref{SETcond}), and therefore rules out
the existence of rotating wormholes for this specific case of
$F_{t\phi}=0$ and $F_{r\theta}=0$.

\section{Conclusion}\label{Conclusion}

In this work we have explored the possibility of the existence of
$(2+1)$ and $(3+1)-$dimensional static, spherically symmetric and
stationary, axisymmetric traversable wormholes coupled to
nonlinear electrodynamics. For the static and spherically
symmetric wormhole spacetimes, we have found the presence of an
event horizon, and that the NEC is not violated at the throat,
proving the non-existence of these exotic geometries within
nonlinear electrodynamics. It is perhaps important to emphasize
that for the $(2+1)-$dimensional case we found an extremely
troublesome aspect of the geometry, as in order to construct a
traversable wormhole, singularities appear in the physical fields.
This particular aspect of the geometry is in clear contradiction
to the model construction of nonlinear electrodynamics, founded on
a principle of finiteness, that a satisfactory theory should avoid
physical quantities becoming infinite \cite{BI}. Thus, imposing
that the physical quantities be non-singular, we verify that the
general solution corresponds to a non-traversable wormhole
geometry. We also point out that the non-existence of
$(3+1)-$dimensional static and spherically symmetric traversable
wormholes is consistent with previous results \cite{Bronnikov1}.

For the $(2+1)-$dimensional stationary and axisymmetric wormhole,
we have verified the presence of an event horizon, rendering a
non-traversable wormhole geometry. Relatively to the
$(3+1)-$dimensional stationary and axially symmetric wormhole
geometry, we have found that the field equations impose specific
conditions that are incompatible with the properties of wormholes.
Thus, we have showed that for the general cases of solutions
outlined above the non-existence of traversable wormholes within
the context of nonlinear electrodynamics. Nevertheless, it is
important to emphasize that regular magnetic time-dependent
traversable wormholes do exist coupled to nonlinear
electrodynamics \cite{Arell-Lobo}.

In the analysis outlined in this paper, we have considered general
relativity coupled to nonlinear electrodynamics, with the
gauge-invariant electromagnetic Lagrangian $L(F)$ depending on a
single invariant $F$ given by $F \sim F^{\mu\nu}F_{\mu\nu}$. An
interesting issue to pursue would be the inclusion, in addition to
$F$, of another electromagnetic field invariant $G \sim
\,^*F^{\mu\nu}F_{\mu\nu}$. This latter inclusion would possibly
add an interesting analysis to the solutions found in this paper.

\section*{Acknowledgements}

We thank Ricardo Garc\'{i}a Salcedo and Nora Bret\'{o}n for
extremely helpful comments and suggestions.


\end{document}